\documentclass[a4paper,twoside]{article}

\usepackage{epsfig}
\usepackage{float}
\usepackage{subcaption}
\usepackage{calc}
\usepackage{amssymb}
\usepackage{amstext}
\usepackage{amsmath}
\usepackage{amsthm}
\usepackage{multicol}
\usepackage{multirow}
\usepackage{pslatex}
\usepackage{url}
\usepackage{algorithm2e}
\usepackage[bottom]{footmisc}
\usepackage{natbib}
\usepackage{SCITEPRESS}     

\begin{document}

\title{MI$^2$DAS: A Multi‑Layer Intrusion Detection Framework with Incremental Learning for Securing Industrial IoT Networks}
\author{\authorname{Wei Lian and Alejandro Guerra-Manzanares\orcidAuthor{0000-0002-3655-5804}\thanks{ \scriptsize{Corresponding author: alejandro.guerra@nottingham.edu.cn}}}
\affiliation{School of Computer Science, University of Nottingham, Ningbo, China}
}


\abstract{The rapid expansion of Industrial IoT (IIoT) systems has amplified security challenges, as heterogeneous devices and dynamic traffic patterns increase exposure to sophisticated and previously unseen cyberattacks. Traditional intrusion detection systems often struggle in such environments due to their reliance on extensive labeled data and limited ability to detect new threats. To address these challenges, we propose MI$^2$DAS, a multi‑layer intrusion detection framework that integrates anomaly‑based hierarchical traffic pooling, open‑set recognition to distinguish between known and unknown attacks and incremental learning for adapting to novel attack types with minimal labeling. Experiments conducted on the Edge‑IIoTset dataset demonstrate strong performance across all layers. In the first layer, GMM achieves superior normal‑attack discrimination (accuracy = 0.953, TPR = 1.000). In open‑set recognition, GMM attains a recall of 0.813 for known attacks, while LOF achieves 0.882 recall for unknown attacks. For fine‑grained classification of known attacks, Random Forest achieves a macro‑F1 of 0.941. Finally, the incremental learning module maintains robust performance when incorporating novel attack classes, achieving a macro-F1 of 0.8995. These results showcase MI$^2$DAS as an effective, scalable and adaptive framework for enhancing IIoT security against evolving threats.}

\onecolumn \maketitle \normalsize \setcounter{footnote}{0} \vfill

\section{\uppercase{Introduction}}
\label{sec:introduction}

The Internet of Things (IoT) stands as a foundational pillar of next-generation information technology, seamlessly connecting the physical and digital worlds through the integration of sensors, embedded systems, wireless communication and cloud computing~\citep{eu_iot_digital_transformation_2025}. At its core, IoT empowers previously isolated devices with environmental awareness, intelligent connectivity and autonomous operational capabilities, reshaping industrial processes and daily life. For instance, \emph{smart} sensors enable real-time monitoring, predictive maintenance and enhanced decision-making across domains from smart cities to healthcare and precision agriculture \citep{rathi2025iot}. The number of IoT devices was estimated at 18.5 billion in 2024 and is projected to grow to 24 billion by 2026, reach 39 billion by 2030 and exceed 50 billion by 2035~\citep{iot_prediction}, spanning a wide range of applications including smart cities, Industry 4.0, healthcare and agricultural monitoring~\citep{Rath2024}, with widespread adoption 
enhancing operational efficiency and resource management.


As a key branch of IoT, the Industrial Internet of Things (IIoT) integrates IoT technologies into industrial environments. Unlike consumer IoT, it emphasizes reliability, real-time performance, controllability and security~\citep{sisinni2018industrial}. By deploying large-scale sensors, actuators, industrial control systems (ICS) and edge computing devices, IIoT facilitates extensive industrial data collection and intelligent analysis, enabling predictive maintenance, automated scheduling and system optimization~\citep{kapoor2025analyzing}. It has seen widespread adoption across manufacturing, energy, smart grids and healthcare, becoming a cornerstone of Industry 4.0 and intelligent manufacturing while improving efficiency, reducing maintenance costs and enhancing operational flexibility~\citep{Angurala2024}. However, the increasing connectivity and openness of IIoT expose industrial infrastructures to growing cybersecurity threats. Unlike traditional IT networks, IIoT devices are resource-constrained, rely on lightweight protocols such as MQTT and Modbus, operate in time-critical environments and have long life cycles with limited update capabilities~\citep{sisinni2018industrial}. These characteristics significantly expand the attack surface and magnify the impact of intrusions, potentially resulting in process manipulation, production disruptions, or large-scale safety incidents, making robust Intrusion Detection Systems (IDS) essential for industrial cybersecurity~\citep{cecilio2024security}.

Existing IDS approaches for IIoT face several significant challenges: (i) struggle with massive, high-dimensional, heterogeneous data streams that show strong temporal dependencies, making real-time detection difficult; (ii) the scarcity of labeled attack samples (especially for emerging, low-frequency attacks) raises the risk of misclassification; severe class imbalance, where normal traffic vastly dominates, degrades the performance of supervised models; and (iii) the rapid emergence of novel, zero-day attacks outpaces traditional signature-based techniques.


To overcome these limitations, a dedicated IIoT intrusion detection architecture is required that enhances the separability between normal and malicious traffic, efficiently handles high-dimensional and imbalanced data and detects novel or zero-day attacks. Achieving these objectives is critical for safeguarding industrial infrastructures, preventing economic losses and enabling the transition towards intelligent, resilient industrial systems. Addressing these needs, our main contributions are: 


\begin{enumerate}
    \item We propose the Multi-layer IIoT Intrusion Detection Adaptive System (MI$^2$DAS) architecture that integrates sequential pooling and classification, enabling accurate separation of normal and attack traffic, effective discrimination between known and unknown threats and incremental discovery and adaptation to emerging attacks.
    \item We perform a comprehensive evaluation of algorithms, identifying complementary strengths and determining the optimal models for each layer within the proposed architecture
    \item We develop an incremental classifier integrating semi-supervised learning or active learning approaches, enabling the continual incorporation of new attack types with minimal labeling effort
    \item We conduct extensive experiments on the Edge-IIoTset dataset to validate the effectiveness and scalability of the proposed architecture. The results demonstrate that the system remains robust under varying attack distributions and across incremental learning stages.
\end{enumerate} 





The structure of this paper is as follows: Section~\ref{sec:related_work} reviews related work on IIoT intrusion detection. Section~\ref{sec:methodology} details the proposed multi-layer intrusion detection framework, MI$^2$DAS. Section~\ref{sec:results} provides the experimental results and analysis, while Section~\ref{sec:limitations} states the limitations of our study. Finally, Section~\ref{sec:conclusions} summarizes the study and outlines future research directions.


\section{\uppercase{RELATED WORK}}
\label{sec:related_work}

IDS have evolved significantly with the advancement of Machine Learning (ML) and Deep Learning (DL) techniques, which enable efficient processing of high-volume network traffic with complex spatiotemporal patterns. Meanwhile, the unique characteristics of the IIoT have driven the development of specialized intrusion detection solutions. This section reviews relevant research on general network intrusion detection algorithms and IIoT-specific IDS.

\subsection{ML-based IDS}

In the context of traditional (non-IoT) networks, ML methods remain widely adopted in intrusion detection due to their efficiency, interpretability and low resource requirements~\citep{ali2025deep}. \citet{Ahmed2022} combined SMOTE oversampling with multi-stage feature selection to optimize Ramdom Forest (RF), achieving 95.1\% accuracy on UNSW-NB15, obtaining better accuracy than models trained on raw imbalanced data. \citet{Kasongo2020} leveraged the gradient boosting model, XGBoost, to select 19 critical features from the 44-dimensional UNSW-NB15 dataset, improving detection accuracy and reducing computational overhead for real-time applications. Although they were only evaluated in non-IoT networks, these optimized pipelines may also benefit resource-constrained IoT systems.

DL methods learn hierarchical features without manual engineering, enabling them to capture complex decision boundaries and significantly improve intrusion detection~\citep{ali2025deep}. Convolutional neural networks (CNN) effectively extract spatial features and, when applied to the CSE-CIC-IDS2018 dataset, achieved 91.5\% accuracy in detecting DoS attacks~\citep{Kim2020}. A hybrid CNN–LSTM architecture, combined with data balancing techniques, achieved over 99\% accuracy on the UNSW-NB15 and CIC-IDS2018 datasets~\citep{AlDener2021}. Addressing labeling challenges, \citet{Shone2018} proposed an autoencoder-based unsupervised feature learning framework followed by RF classification, achieving good results on NSL-KDD.

\subsection{Intrusion Detection in IIoT}

IIoT environments, which integrate ICS, sensor networks and cloud/edge infrastructures, face significant challenges such as protocol heterogeneity (e.g., Modbus, CAN, MQTT), strict latency demands, resource-limited edge devices and evolving attack surfaces~\citep{sisinni2018industrial}. Optimized ML pipelines are widely adopted in IIoT environments due to their low computational overhead. In this regard, \citet{Kasongo2020} used XGBoost-based feature selection and achieved real-time detection with latency lower than 50ms on IIoT-like traffic, while \citet{Talukder2024} combined random oversampling and stacking feature embedding to improve performance on IIoT-related attacks, reaching 99.95\% accuracy on CIC-IDS2017 and reducing false positives by 62\%. \citet{Mohy2022} proposed a lightweight hybrid model combining Pearson's correlation coefficient, Isolation Forest (IF) and RF, which achieved 98.3\% accuracy on BoT-IoT. 

Hybrid DNN/CNN–LSTM frameworks tailored for protocols such as Modbus/TCP can extract protocol-aware features and sequential control command patterns, achieving high DoS detection accuracy with low inference latencies~\citep{halbouni2022cnn}. Similarly, protocol-specific approaches, such as \citet{Deng2022} voltage fingerprint-based IDS for CAN bus networks leverage physical-layer characteristics to reach 99.2\% with $<1$ ms latency. \citet{Lo2022} proposed E-GraphSAGE, the first Graph Neural Network-based IIoT IDS that integrates edge features and network topology to capture collaborative malicious behaviors among compromised devices, achieving F1-scores up to 1.0 on BoT-IoT. \citet{Saez2023} proposed a privacy-preserving architecture to train unsupervised models for network intrusion detection in large, distributed IoT and IIoT deployments.

Although existing methods address specific challenges within IIoT networks, our proposed approach, MI$^2$DAS, tackles the core challenges in IIoT environments by employing lightweight models, efficiently managing high-dimensional and imbalanced data and enabling effective detection of novel attack types. By incorporating new attacks into the detection pipeline via semi-supervised and active learning strategies, MI$^2$DAS not only maintains high performance, but also reduces the need for extensive manual labeling.

\section{\uppercase{METHODOLOGY}}
\label{sec:methodology}
The proposed methodology introduces a multi-layer intrusion detection architecture tailored for IIoT environments, addressing three core security challenges:

\begin{itemize}
    \item \emph{Intrusion detection}. The proposed system effectively distinguishes normal traffic from malicious activity at the network edge.
    \item \emph{Novel attack detection}. The architecture enables the identification of previously unseen or zero-day attacks.
    \item \emph{Adaptive continuous learning}. The system maintains detection performance in the presence of evolving attack types by incrementally incorporating new threats under limited labeling resources.
\end{itemize}

The proposed architecture is organized into three sequential layers, each addressing a distinct functional requirement:

\begin{itemize}
    \item \emph{Layer 1: Traffic Filtering and Normal Flow Processing}. Performs the initial binary separation of traffic, distinguishing benign flows from suspicious ones at the network edge.
    \item \emph{Layer 2: Novelty Detection and Known Attack Classification}. Identifies previously unseen attack patterns while classifying traffic into established attack categories at the network edge. 
    \item \emph{Layer 3: Incremental Learning and Adaptive Modeling}. Unseen attack patterns are relayed to the central server for analysis by the edge nodes. This layer incrementally incorporates new attack types into the detection pipeline, enabling the system to maintain performance under evolving threats.
\end{itemize}

\begin{figure*}[h]
    \centering
    \includegraphics[width=\textwidth]{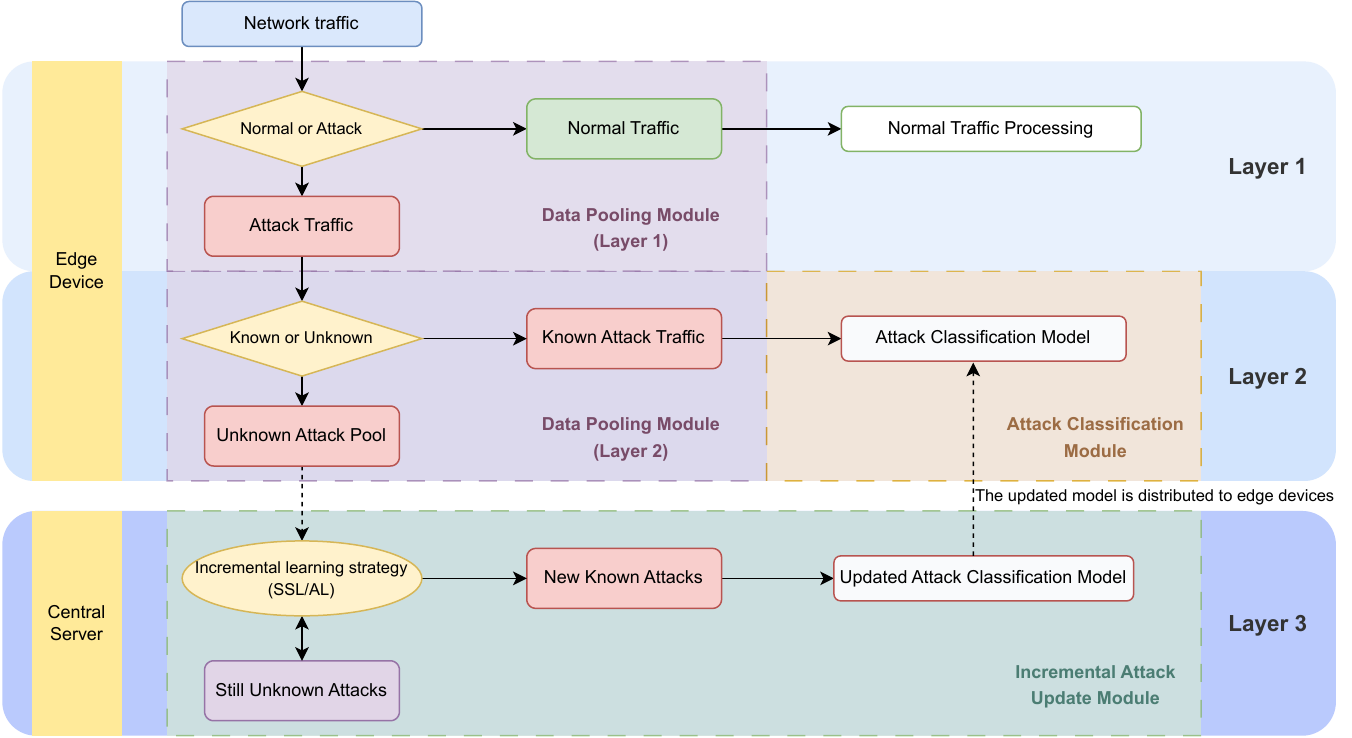}
    \caption{Architecture of the Multi-Layer IIoT Intrusion Detection Adaptive System (MI$^2$DAS).}
    \label{fig:miidas}
\end{figure*}

Figure~\ref{fig:miidas} illustrates the architectural design of the proposed system. The hierarchical structure enables IIoT devices to perform lightweight, early-stage traffic filtering (Layer 1: normal vs. attack traffic separation; Layer 2: known vs. unknown attack separation and known attack classification), while the server-side model performs progressively refined traffic analysis and dynamically updates the detection capabilities of the system to evolving threats.

As depicted in Figure~\ref{fig:miidas}, the MI$^2$DAS architecture is composed of three core modules, outlined as follows and elaborated in the subsequent subsections:

\begin{itemize}
    \item \emph{Data Pooling Module}. This module comprises two layers, which are deployed on edge devices. It performs traffic filtering at multiple levels of granularity, organizing flows into distinct data pools to enable efficient subsequent analysis. The two‑level module is enclosed in a purple box in Figure~\ref{fig:miidas}.
    
    \item \emph{Attack Classification Module}. This edge-device component provides a fine-grained categorization of known attack types, enabling accurate identification and differentiation of malicious traffic patterns (attacks). This layer 2 module is highlighted in an orange box in Figure~\ref{fig:miidas}.
    
    \item \emph{Incremental Attack Update Module}. This server-side component is designed to maintain detection performance under evolving threats. This module incrementally incorporates newly emerging attack types using semi-supervised and active learning strategies, reducing the need of extensive manual annotation. Figure~\ref{fig:miidas} highlights this layer 3 module within a green box.
\end{itemize}


\subsection{Data Pooling Module}

The Data Pooling Module (DPM), is composed of two layers deployed on edge devices. It performs a two-stage hierarchical partitioning of raw network traffic, classifying incoming data into different data pools. It adopts novelty detection and outlier detection models to minimize dependency on labeled attack samples and to ensure system functionality under real-world class-imbalance scenarios. 

\subsubsection{First DPM Layer: Normal vs. Attack}
\label{sec:first-layer-dpm}
The first-level detector separates benign traffic from anomalous patterns using novelty detection or outlier detection techniques. Since only normal traffic is guaranteed to be available at deployment time, the system leverages unsupervised boundary-learning models, such as~\citep{Chandola2009}:

\begin{enumerate}
    \item \emph{One-Class Support Vector Machines} (OC-SVM): It constructs a decision boundary in a high-dimensional feature space around normal traffic using a maximum-margin formulation. This approach is effective for modeling compact normal patterns, but its performance is highly sensitive to the choice of kernel function.

    \item \emph{Gaussian Mixture Models} (GMM): It models data as a weighted mixture of Gaussian components and performs anomaly detection by evaluating the likelihood of a sample under the estimated density. Its probabilistic and multi-modal representation provides flexibility for capturing heterogeneous normal behaviors.

    \item \emph{Local Outlier Factor} (LOF): It quantifies the local density deviation of each sample relative to its neighbors, assigning higher anomaly scores to points that are locally sparse or isolated. As a non-parametric, density-based approach, it is particularly effective at capturing context-dependent and irregular anomalies.
\end{enumerate}

Note that, depending on the characteristics of the data traffic, different anomaly detection models may show varying levels of effectiveness. In this regard, density-based approaches, such as LOF, are well-suited for capturing local irregularities, while probabilistic models like GMM are more effective at representing heterogeneous traffic distributions. Boundary-learning methods such as OC-SVM perform best when normal traffic patterns are compact and well-defined.

The detection model is trained offline using only normal samples from the central server and then deployed to edge devices to execute real-time, per-sample traffic pooling.

\subsubsection{Second DPM Layer: Open-Set Recognition for Attack Categorization}
\label{sec:second-layer-dpm}

As shown in Figure~\ref{fig:miidas}, traffic identified as abnormal in the first layer is further processed to distinguish:

\begin{itemize}
    \item \emph{Known attack categories}, which are classified using supervised models trained on labeled attack samples.
    \item \emph{Novel or unknown attack patterns}, which are flagged through open-set recognition techniques and forwarded for adaptive learning.
\end{itemize}

This stage again relies on novelty detection since only known attack categories are available during training. To accommodate diverse deployment scenarios, in our experiments, multiple attack-category partitions are considered, allowing edge devices to operate flexibly under heterogeneous attack distributions. In our experiments, the models evaluated at this stage include OC-SVM, GMM, LOF and IF. As depicted in Figure~\ref{fig:miidas}, the output of this layer is organized into a two-pool structure: a \emph{Known Attack Traffic} pool and an \emph{Unknown Attack Pool}, which support downstream classification tasks (Section \ref{sec:data_classification_module}) and facilitate incremental model updates (Section \ref{sec:iterative_data_classification_module}), respectively.

\subsection{Attack Classification Module}
\label{sec:data_classification_module}

The Attack Classification Module processes traffic assigned to the \emph{Known Attack Traffic} pool (i.e., the output of the second DPM layer as described in Section~\ref{sec:second-layer-dpm}), performing fine-grained traffic classification in established attack categories. This component leverages supervised learning models specifically optimized for high-dimensional IIoT traffic, ensuring accurate differentiation among attack types.

In our experimental setup, we evaluated two types of classification models:

\begin{enumerate}
    \item \emph{Traditional ML Models}. k-Nearest Neighbors (k-NN), Support Vector Machines (SVM) and Logistic Regression (LR) were used as baseline classifiers. These models serve as benchmarks and provide interpretable decision boundaries.
    \item \emph{Ensemble Models}. To capture non-linear patterns and improve robustness, the following ensemble methods were adopted as core classifiers:
    \begin{itemize}
        \item \emph{RF}: Bootstrap-aggregated decision trees that ensure stability and resistance to noise.
        \item \emph{XGBoost}: Gradient boosting trees with second-order optimization and built-in regularization.
        \item \emph{LightGBM}: Gradient boosting trees optimized for efficiency and scalability, incorporating second-order optimization and regularization.
    \end{itemize}
\end{enumerate}

Similar to the anomaly detection models described in Section~\ref{sec:first-layer-dpm}, these classifiers are trained at the central server on structured attack features and subsequently deployed to edge devices to efficiently classify traffic within the \emph{Known Attack Pool}.

\subsection{Incremental Attack Update Module}
\label{sec:iterative_data_classification_module}

The Incremental Attack Update Module ensures long-term adaptability by incorporating new attack types, using data assigned to the \emph{Unknown Attack Pool} identified by the second DPM layer. as outlined in Section~\ref{sec:second-layer-dpm}. Through continual learning mechanisms, the taxonomy is incrementally expanded to incorporate novel attack classes, while mitigating catastrophic forgetting to preserve performance on previously learned categories.

Note that since labeled data for new attack classes is often unavailable, the central server employs incremental methods to further differentiate:

\begin{itemize}
    \item \emph{New attack types}, which are incorporated into the system through incremental class expansion.
    \item \emph{Still unknown attacks}, which remain flagged for further analysis, labeling, or adaptive retraining in subsequent cycles.
\end{itemize}

Based on the operational characteristics of IIoT environments, we propose two incremental learning strategies for incremental learning: Semi-Supervised Learning (SSL) and Active Learning (AL). They are briefly described as follows:

\begin{enumerate}
    \item \emph{Semi-Supervised Learning}. This approach automatically expands the labeled dataset by assigning pseudo-labels to high-confidence unknown samples. Key algorithms include: self-training, label propagation and label spreading. This approach leverages the abundance of unlabeled data to improve model generalization with minimal manual effort.
    \item \emph{Active Learning}. This strategy selects the most informative samples, using uncertainty-based or representativeness-based criteria, for manual expert annotation, minimizing the labeling effort while maximizing the utility of newly labeled data.
\end{enumerate}

These strategies enables adaptive, incremental expansion of the attack taxonomy with minimal labeling efforts, avoiding full dataset retraining while preserving scalability and responsiveness in dynamic threat environments.

The detailed incremental training procedure is presented in pseudocode in Algorithm~\ref{alg:1}. In our experimental setup, we evaluated three commonly used SSL approaches: self-training, label spreading and label propagation. For active learning, we used uncertainty-based sampling strategies to query the most informative instances for manual annotation. Note that, for compactness, we include both SSL and AL approaches within Algorithm~\ref{alg:1}, although in practice they would typically be employed independently depending on the availability of unlabeled data and labeling resources (i.e., human experts, often referred to as \emph{oracles} in the AL nomenclature).

\begin{algorithm}[!h]
\caption{Adaptive model training via SSL and AL}
\label{alg:1}
\KwData{Known Attack Pool $KnownAttack$, Unknown Attack Pool $UnknownPool$}
\KwResult{Updated classifier $C$ and refined $UnknownPool$}

Initialize seed set $S$ from manually annotated samples\;
$L \leftarrow KnownAttack \cup S$\;
Initialize base classifier $C$\;
Initialize semi-supervised learner $SSL$ (i.e., self-training, label spreading or label propagation method) or active learner $AL$\;

\For{each iteration}{
    \uIf{Strategy = ``SSL''}{
        Generate pseudo-labels for subset $U \subseteq UnknownPool$ using $SSL$\;
        Select high-confidence pseudo-labeled set $P$\;

        \uIf{UsePseudoLabeling = TRUE}{
            $L \leftarrow L \cup P$\;
            $UnknownPool \leftarrow UnknownPool \setminus P$\;
        }
        \Else{
            Retain $P$ for monitoring or validation only\;
        }
    }
    \ElseIf{Strategy = ``AL''}{
        Select most informative samples $Q \subseteq UnknownPool$ using $AL$ (e.g., uncertainty sampling)\;
        Query ground-truth labels for $Q$ from human oracle\;
        $L \leftarrow L \cup Q$\;
        $UnknownPool \leftarrow UnknownPool \setminus Q$\;
    }

    Retrain classifier $C$ on updated labeled set $L$\;

    \If{convergence criterion satisfied}{
        \textbf{break}\;
    }
}
\end{algorithm}

\subsection{Summary of Methodological Advantages}
This study introduces MI$^2$DAS, a multi-level intrusion detection architecture specifically designed for IIoT environments. The proposed framework addresses key challenges inherent to IIoT environments, including limited computational resources, scarcity of labeled training data, class imbalance and the continuous emergence of novel threats. 

The architecture is organized as a hierarchical collaboration of three core modules:

\begin{itemize}
    \item \textbf{Data Pooling}: Aggregates heterogeneous IIoT traffic into structured pools, facilitating hierarchical organization and enabling efficient separation of normal and malicious flows at multiple levels of granularity.
    \item \textbf{Attack Classification}: Employs supervised learning models optimized for high-dimensional IIoT traffic to achieve fine-grained categorization of known attack types.
    \item \textbf{Incremental Attack Update}: Ensures long-term adaptability through continual learning, dynamically integrating newly discovered attack classes while mitigating catastrophic forgetting.
\end{itemize}

Integrated within a unified pipeline, as depicted in Figure~\ref{fig:miidas}, the modules enable continual learning and adaptive intrusion detection, ensuring efficient operation while preserving robustness against the dynamic threat landscape of IIoT systems.


\section{\uppercase{RESULTS \& DISCUSSION}}
\label{sec:results}

This section presents a comprehensive evaluation of the proposed multi-layer intrusion detection and continual learning pipeline. To validate the effectiveness of the architecture, four experiments are conducted, each corresponding to one of the operational stages of the system. For every experiment, we detail the experimental design, report the results and provide in-depth discussion, with supporting tables and figures referenced where appropriate.

\subsection{Dataset}
The experiments are based on the Edge-IIoTset dataset~\citep{Ferrag2022}, a large-scale industrial IoT traffic dataset specifically designed for evaluating intrusion detection systems for IIoT networks. Specifically, it contains:

\begin{itemize}
    \item \textbf{Normal Traffic}: Generated by benign industrial control system operations, representing baseline IIoT behavior.
    \item \textbf{Attack Traffic}: Fourteen distinct classes, including Distributed Denial-of-Service (DDoS) attacks (TCP/UDP/ICMP/HTTP), password-based attacks, backdoor intrusions, Man-In-The-Middle (MITM), SQL injection, fingerprinting, vulnerability scanning, cross-site scripting (XSS), malicious file uploading and other threats.
    \item \textbf{Feature Space}: High-dimensional and heterogeneous, extracted from both network flow statistics and IoT protocol-level interactions. To avoid overfitting, we excluded eight topology‑dependent features (e.g., IPs), resulting in a final set of 53 features.
    \item \textbf{Class Imbalance}: The dataset has a significantly skewed data distribution, with normal traffic making up most of the data while some attack categories have only a small number of samples.
\end{itemize}

This diversity and imbalance make Edge-IIoTset an appropriate benchmark for simulating realistic IIoT environments, where normal traffic constitutes the majority, novel attacks emerge continuously and labeled samples are scarce. Note that while the complete dataset comprises $\approx 20$ million IIoT traffic records, for reproducibility, we rely on the official dataset splits. Specifically, Table~\ref{tab:edgeiiot} provides the class distribution of the randomly selected subsets of data for ML algorithms from the Edge-IIoTset dataset according to the official training-test partition provided by the authors~\citep{Ferrag2022}.


\subsection{DPM Layer 1: Normal or Attack}

The first DPM layer serves as the system’s initial defense layer, designed to filter malicious traffic from legitimate and normal activity. All experiments use the Edge-IIoTset dataset, which includes normal traffic and fourteen different attack types, as described in Table~\ref{tab:edgeiiot}.

\begin{table}[h]
\caption{Dataset class distribution~\citep{Ferrag2022}}
\label{tab:edgeiiot}
\centering
\scriptsize
\begin{tabular}{|l|c|c|c|}
\hline
Class & Total & Training & Test \\
\hline
Normal & 24301 & 19281 & 4820  \\ \hline
Backdoor & 10195 & 7892 & 1973 \\
DDoS\_HTTP & 10561 & 8396 & 2099  \\
DDoS\_ICMP & 14090 & 10477 & 2619  \\
DDoS\_TCP & 10247 & 8198 & 2049  \\
DDoS\_UDP & 14498 & 11598 & 2900  \\
Fingerprinting & 1001 & 682 & 171 \\
MITM & 1214 & 286 & 72 \\
Password & 9989 & 7978 & 1994  \\
Port Scanning & 10071 & 7137 & 1784 \\
Ransomware & 10925 & 7751 & 1938 \\
SQL Injection & 10311 & 8225 & 2057 \\
Uploading & 10269 & 8171 & 2043 \\
Vulnerability Scan & 10076 & 8050 & 2012 \\
XSS attack & 10052 & 7634 & 1909 \\
\hline
\end{tabular}
\end{table}

In this experiment, two anomaly detection paradigms are considered: \emph{Novelty detection}, which trains the model exclusively on normal traffic under the assumption that anomalies (attacks) are absent during training, thus modeling the challenge of zero-day attack detection; and \emph{outlier detection}, which trains the model on predominantly normal traffic with a minor proportion of attack samples (100:1 ratio), assuming contamination in the training set and simulating lightly contaminated real-world environments.

Two balanced test sets are constructed: one comprising 1,000 normal and 1,000 attack samples and another comprising 5,000 normal and 5,000 attack samples. At this layer, the models evaluated include OC-SVM, GMM and LOF. Model performance is assessed using accuracy, True Positive Rate (TPR, also known as Recall), False Positive Rate (FPR) and precision. We conduct multiple iterations for each model using different hyperparameter configurations to ensure robust performance evaluation.

\subsubsection{Experiment Results}

In our experiments, GMM consistently yields the highest performance across both novelty and outlier detection settings, highlighting its strong ability to model complex traffic distributions and adapt to varying contamination levels. OC-SVM achieves perfect recall in the novelty setting, demonstrating its effectiveness in detecting all attacks when trained on clean data. However, it produces a high number of false alarms (FPR 0.410), which limits its practical applicability in this scenario. LOF shows unstable performance with significant variance across test subsets, likely due to its reliance on local density estimation, which is sensitive to heterogeneous and high-dimensional IIoT traffic. The top three GMM results for each detection setting are presented in Table~\ref{tab:table2}.

\begin{table}[h]
\caption{Top-3 GMM performance per setting. The parameter \emph{nc} denotes the number of mixture components, while \emph{th\_per} indicates the threshold percentile used for anomaly classification. The best results are highlighted in bold.}
\label{tab:table2}
\centering
\resizebox{\columnwidth}{!}{
\begin{tabular}{|c|c|c|c|c|c|}
\hline
\textbf{Setting} & \textbf{Parameter} & \textbf{Acc.} & \textbf{TPR} & \textbf{FPR} & \textbf{Pr.} \\
\hline

\multirow{3}{*}{Novelty}
 & \textbf{nc=2, th\_per=5} & \textbf{0.953} & \textbf{1.000} & \textbf{0.095} & \textbf{0.914} \\
 & nc=4, th\_per=5 & 0.950 & 1.000 & 0.099 & 0.910 \\
 & nc=3, th\_per=5 & 0.950 & 1.000 & 0.100 & 0.909 \\
\hline

\multirow{3}{*}{Outlier}
 & nc=2, th\_per=5 & 0.944 & 1.000 & 0.112 & 0.899 \\
 & nc=4, th\_per=5 & 0.810 & 1.000 & 0.381 & 0.724 \\
 & nc=3, th\_per=5 & 0.804 & 1.000 & 0.392 & 0.718 \\
\hline

\end{tabular}
}
\end{table}

\subsubsection{Discussion}

The first DPM layer requires extremely high recall to prevent attacks from passing to subsequent stages. At the same time, controlling the FPR is critical to minimize false alarms and prevent misclassification of a large proportion of normal traffic. Among the evaluated models, GMM achieves a superior balance of high recall and low FPR, making it the most effective choice for this layer. These results demonstrate that a probabilistic density-based approach is best suited for the heterogeneous and high-variance IIoT traffic patterns observed in the Edge-IIoTset dataset, where accurate modeling of complex distributions is essential for robust anomaly detection.

\subsection{DPM Layer 2: Open-Set Recognition for Attack Categorization}

The second DPM layer further refines the filtering process by separating attack traffic (output from the first DPM layer) into two categories: \emph{known attack types}, which are used for supervised learning later and \emph{unknown attack types}, which are isolated to prevent unseen attack behaviors from contaminating the supervised classifier.

To assess the robustness of our approach, we designed five experimental configurations using the EdgeIIoT dataset. Each configuration varies the proportion of known and unknown attack types to simulate different levels of uncertainty in real-world scenarios. Specifically, the configurations are: (i) 1 known attack type and 13 unknown types, (ii) 4 known and 10 unknown, (iii) 7 known and 7 unknown, (iv) 10 known and 4 unknown and (v) 13 known and 1 unknown. This progressive variation allows us to evaluate how the system performs when the classifier has minimal prior knowledge versus when it has extensive prior knowledge of attack behaviors. We evaluate four anomaly detection models, i.e., GMM, LOF, OC-SVM and IF. For each configuration, we systematically evaluate all possible combinations of known and unknown attack types. For example, in the configuration with 1 known and 13 unknown attacks, there are 14 possible combinations (choosing which single attack is known). In contrast, the configuration with 7 known and 7 unknown attacks yields 3,432 possible combinations. This exhaustive evaluation ensures that our results are not biased by a specific selection of attack types and provides a comprehensive assessment of model performance under varying knowledge distributions.


\subsubsection{Experiment Results}

\begin{figure*}[h]
    \centering
    \includegraphics[width=\textwidth]{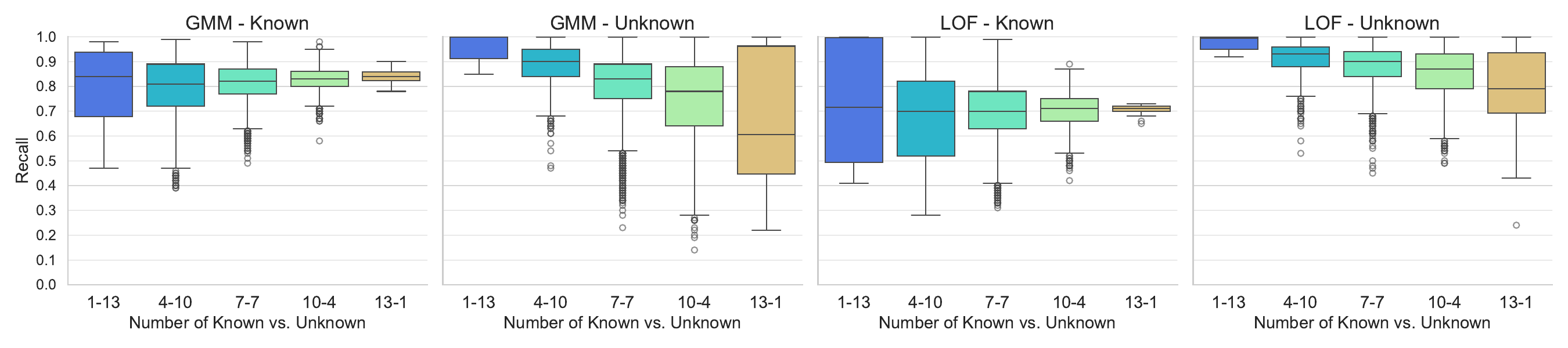}
    \caption{Recall performance for GMM and LOF methods for GMM and LOF methods}
    \label{fig:classifier_results}
\end{figure*}

In our experiments, across all configurations, GMM and LOF consistently outperform OC-SVM and IF. Figure~\ref{fig:classifier_results} presents boxplots summarizing the performance across all possible combinations of known and unknown attack types. Figure~\ref{fig:classifier_results} also highlights performance biases in LOF and GMM, suggesting that these models tend to favor specific distributions of known and unknown attacks, which may influence their generalization capability.




\subsubsection{Discussion}

In the second DPM layer, GMM and LOF show complementary strengths in recognizing known and unknown attack types. As shown in Figure~\ref{fig:classifier_results}, GMM achieves strong recall for known attacks, with an average of 0.813±0.086, whereas LOF demonstrates superior recall for unknown attacks, averaging 0.882±0.080. These results indicate that density-based (LOF) and probabilistic (GMM) approaches are better suited to capture the complex distribution of IIoT attack behaviors compared to margin-based (OC-SVM) and tree-based (IF) methods. The complementarity between GMM and LOF suggests that probabilistic modeling particularly effective when attack patterns are well-represented in training data, whereas LOF's density-based approach is more effective in detecting deviations from the training data, making it particularly suitable for identifying novel threats. This highlights the potential of hybrid strategies that combine both methods to achieve balanced detection performance across both known and unknown attack spaces.

\subsection{Attack Classification Module}

The attack classification model in the ACM module processes \emph{Known Attack Traffic} and assigns each sample to its corresponding attack category, performing multiclass classification.

In our experimental setup, we evaluate six classification methods (as described in Section~\ref{sec:data_classification_module}) under multiple configurations. Specifically, we perform 50 random combinations of known attack sets for each scenario involving 4 Known, 7 Known and 10 Known classes. Additionally, we evaluate all 14 possible combinations for the set of 13 known attacks scenario, where only one attack class is excluded each time. Classification performance is evaluated using micro accuracy, macro accuracy, macro F1 and weighted F1 scores.



\subsubsection{Experiment Results}

\begin{figure*}[h]
    \centering
    \includegraphics[width=\textwidth]{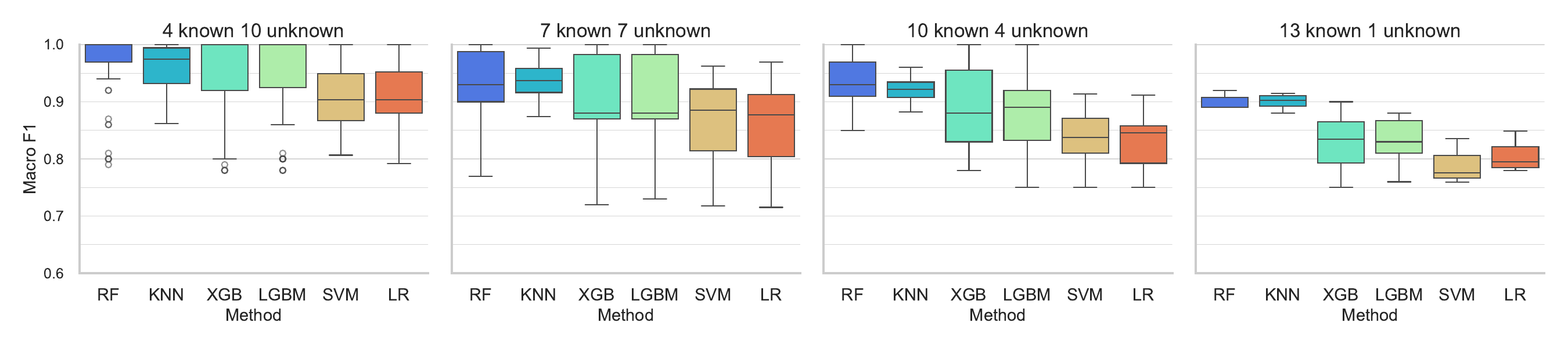}
    \caption{Macro F1 performance values for different classification models and Known-Unknown combinations}
    \label{fig:figure_3}
\end{figure*}

The boxplots in Figure~\ref{fig:figure_3} illustrate the distribution of macro‑F1 performance across different classification algorithms and varying combinations of known and unknown attacks, while Table~\ref{tab:table3} provides the aggregated results for all 164 combinations. Specifically, these include 50 random combinations for the 4‑Known, 7‑Known and 10‑Known attack settings, as well as the 14 possible combinations for the complete set of 13 known attacks.

\begin{table}[h]
\caption{Aggregated results across all combinations and evaluated algorithms. The best results are highlighted in bold.}
\label{tab:table3}
\centering
\resizebox{\columnwidth}{!}{
\begin{tabular}{|c|c|c|c|c|}
\hline
Model  & Macro-F1 & Weighted-F1 & Macro-Accuracy & Micro-Accuracy \\
\hline
RF & \textbf{0.941 $\pm$ 0.054} & \textbf{0.944 $\pm$ 0.056} & \textbf{0.950 $\pm$ 0.048} & \textbf{0.949 $\pm$ 0.050}  \\
KNN  & 0.938 $\pm$ 0.033 & 0.938 $\pm$ 0.036 & 0.937 $\pm$ 0.034 & 0.938 $\pm$ 0.040 \\
XGBoost  & 0.906 $\pm$ 0.079 & 0.910 $\pm$ 0.083 & 0.924 $\pm$ 0.068 & 0.920 $\pm$ 0.072 \\
LightGBM  & 0.905 $\pm$ 0.079 & 0.908 $\pm$ 0.083 & 0.922 $\pm$ 0.068 & 0.918 $\pm$ 0.073 \\
SVM  & 0.866 $\pm$ 0.064 & 0.883 $\pm$ 0.063 & 0.882 $\pm$ 0.059 & 0.885 $\pm$ 0.060 \\
LR  & 0.862 $\pm$ 0.063 & 0.881 $\pm$ 0.061 & 0.887 $\pm$ 0.057 & 0.882 $\pm$ 0.059 \\
\hline
\end{tabular}
}
\end{table}

\subsubsection{Discussion}

The results indicate that, given sufficient training data, both traditional supervised learning methods and ensemble approaches achieve high classification performance. Among these, RF consistently outperforms all other models, achieving scores above 0.941 across all evaluated metrics. The second-best performer is k-NN, which demonstrates stable results with an average of 0.938 across metrics. In contrast, linear models such as SVM and LR report the lowest performance, with all scores falling below 0.887. 

These results are consistent with prior work in the IoT domain \citep{guerra2020using}, supporting the effectiveness of RF in handling high-dimensional, heterogeneous IIoT traffic. Its inherent robustness to class imbalance, ability to capture complex non-linear decision boundaries and resistance to overfitting make RF particularly well-suited for this domain. Furthermore, RF offers practical advantages such as interpretability and scalability, which are essential for real-world deployments where adaptability and transparency are critical. These features make RF an ideal candidate for the attack classification layer of the system, ensuring reliable detection even in dynamic and evolving threat landscapes. Overall, these findings highlight the superiority of tree-based ensemble methods and instance-based methods in modeling complex relationships within IIoT traffic, while linear models struggle to achieve comparable performance in these conditions.

\subsection{Incremental Attack Update Module}

The purpose of the Incremental Attack Update Module is to maintain the detection performance of the attack classification model within the dynamic IIoT threat landscape by periodically updating the classifier knowledge with new attack categories. Therefore, in this experiment, we evaluate the system’s capability to continuously incorporate new attack categories. To achieve this, we consider two incremental learning schemes (i.e., one-step and multiple-step) and compare the performance of different update strategies (i.e., semi-supervised methods and active learning). A detailed description of the experimental setup is provided as follows.

\begin{enumerate}
    \item \emph{One-Step Iteration}: In this scheme, the update process transitions from $N$ \emph{Known} attacks and $14 - N$ \emph{Unknown} attacks to $14$ \emph{Known} and $0$ \emph{Unknown} attacks. That is, the model starts with knowledge of $N$ attacks and aims to integrate all remaining attacks in a single update step. 
    
    In our experiments, we consider $N = \{4, 7, 10, 13\}$ and evaluate two strategies: \emph{semi-supervised learning} and \emph{active learning}, across five random configurations (i.e., selecting the $N$ attacks at random) for each run. We evaluate three semi-supervised methods (i.e., self-training, label propgation and label spreading) and uncertainty-sampling-based active learning.
    \item \emph{Mult-step Iteration}: This scheme defines a progressive update process, which simulates the progressive emergence of new attack types as follows:  \emph{4K \& 10U $\rightarrow$ 7K \& 7U $\rightarrow$ 10K \& 4U $\rightarrow$ 13K \& 1U $\rightarrow$ 14K \& 0U}, where \emph{K} refers to Known attacks and \emph{U} to Unknown attacks. In this case, the model is progressively refined through multiple incremental steps, starting from 4 Known and 10 Umknown attacks and ending at 14 Known and 0 Unknown attacks, as in the \emph{One-Step Iteration scheme}. 
    As in the previous scheme, we evaluate two strategies: \emph{semi-supervised learning} and \emph{active learning}, across five random configurations of the initial Known attacks. In addition, for the multi-step updates, we compare two training logics for handling pseudo-labeled data:
\begin{enumerate}
    \item \emph{Strict Seed-Based Training}: Using only the original labeled seed samples in each iteration.
    \item \emph{Incremental Augmentation}: Incorporating pseudo-labeled data into subsequent iterations.
\end{enumerate}
\end{enumerate}

\subsubsection{Experiment Result}

Table~\ref{tab:table4} shows the performance of the evaluated methods for $N = 4$ Known attacks, while Table~\ref{tab:table5} reports the same information for $N = 7$ Known attacks. Similarly, Table~\ref{tab:table6} shows the performance of the evaluated methods for $N = 10$ Known attacks, while Table~\ref{tab:table7} reports the same information for $N = 13$ Known attacks. In all cases, the corresponding Unknown attacks is $14 - N$ and the update process is one step, that is, all new unknown attacks are discovered at once and present at the \emph{Unknown attack pool}. 

Table~\ref{tab:multi_step} provides the results of the multi‑step iteration procedure under two training logics for pseudo‑labeled data: one that strictly relies on seed samples (i.e., seed-based in the table) and another that incorporates pseudo‑labeled data into subsequent iterations (i.e., augmentation). The reported results are averages of 5 random runs. 

\begin{table}[H]
\renewcommand{\arraystretch}{0.88}
\caption{Average performance of One-Step Iteration for $N = 4$ Known Attacks (5 random runs). The best results are highlighted in bold.}
\label{tab:table4}
\centering
\resizebox{\columnwidth}{!}{
\begin{tabular}{|c|c|c|c|}
\hline
Method &  Macro F1 & Balanced Acc & Acc \\
\hline
Self-training  & \textbf{0.8859 $\pm$ 0.0182} & \textbf{0.8954 $\pm$ 0.0183} & \textbf{0.8995 $\pm$ 0.0140} \\
Label Spreading  & 0.8349 $\pm$ 0.0337 & 0.8561 $\pm$ 0.0258 & 0.8590 $\pm$ 0.0354 \\
Label Propagation  & 0.8404 $\pm$ 0.0210 & 0.8632 $\pm$ 0.0227 & 0.8630 $\pm$ 0.0186 \\
Active Learning  & 0.8296 $\pm$ 0.0353 & 0.8561 $\pm$ 0.0215 & 0.8548 $\pm$ 0.0254 \\
\hline
\end{tabular}
}
\end{table}

\begin{table}[H]
\renewcommand{\arraystretch}{0.88}
\caption{Average performance of One-Step Iteration for $N = 7$ Known Attacks (5 random runs). The best results are highlighted in bold.}
\label{tab:table5}
\centering
\resizebox{\columnwidth}{!}{
\begin{tabular}{|c|c|c|c|}
\hline
Method  & Macro F1 & Balanced Acc & Acc \\
\hline
Self-training & \textbf{0.8648 $\pm$ 0.0090} & \textbf{0.8908 $\pm$ 0.0127} & \textbf{0.8834 $\pm$ 0.0100} \\
Label Spreading  & 0.7788 $\pm$ 0.0299 & 0.8188 $\pm$ 0.0274 & 0.8065 $\pm$ 0.0334 \\
Label Propagation  & 0.8007 $\pm$ 0.0373 & 0.8350 $\pm$ 0.0326 & 0.8241 $\pm$ 0.0379 \\
Active Learning& 0.8113 $\pm$ 0.0240 & 0.8553 $\pm$ 0.0144 & 0.8504 $\pm$ 0.0174  \\
\hline
\end{tabular}
}
\end{table}

\begin{table}[H]
\renewcommand{\arraystretch}{0.88}
\caption{Average performance of One-Step Iteration for $N = 10$ Known Attacks (5 random runs). The best results are highlighted in bold.}
\label{tab:table6}
\centering
\resizebox{\columnwidth}{!}{
\begin{tabular}{|c|c|c|c|}
\hline
Method  & Macro F1 & Balanced Acc & Acc \\
\hline
Self-training  & \textbf{ 0.8617 $\pm$ 0.0102} & \textbf{0.8835 $\pm$ 0.0119} & \textbf{0.8822 $\pm$ 0.0076} \\
Label Spreading & 0.7753 $\pm$ 0.0554 & 0.8044 $\pm$ 0.0424 & 0.8042 $\pm$ 0.0472 \\
Label Propagation & 0.8019 $\pm$ 0.0426 & 0.8270 $\pm$ 0.0404 & 0.8280 $\pm$ 0.0379 \\
Active Learning  & 0.8303 $\pm$ 0.0185 & 0.8664 $\pm$ 0.0221 & 0.8655 $\pm$ 0.0224 \\
\hline
\end{tabular}
}
\end{table}

\begin{table}[H]
\renewcommand{\arraystretch}{0.88}
\caption{Average performance of One-Step Iteration for $N = 13$ Known Attacks (5 random runs). The best results are highlighted in bold.}
\label{tab:table7}
\centering
\resizebox{\columnwidth}{!}{
\begin{tabular}{|c|c|c|c|}
\hline
Method  & Macro F1 & Balanced Ac c& Acc \\
\hline
Self-training & \textbf{0.8970 $\pm$ 0.0089} & \textbf{0.9023 $\pm$ 0.0061} & \textbf{0.9053 $\pm$ 0.0066} \\
Label Spreading  & 0.8508 $\pm$ 0.0382 & 0.8752 $\pm$ 0.0341 & 0.8692 $\pm$ 0.0371 \\
Label Propagation  & 0.8590 $\pm$ 0.0219 & 0.8836 $\pm$ 0.0206 & 0.8765 $\pm$ 0.0222 \\
Active Learning  & 0.8128 $\pm$ 0.0169 & 0.8377 $\pm$ 0.0113 & 0.8367 $\pm$ 0.0119 \\
\hline
\end{tabular}
}
\end{table}

\begin{table}[htbp]
\centering
\caption{Performance results of multi-step iteration for different strategies. The best results per step are highlighted in bold.}
\label{tab:multi_step}
\resizebox{\columnwidth}{!}{
\begin{tabular}{|l|l|c|c|c|}
\hline
Step & Strategy & Macro-F1 & Balanced Acc & Accuracy  \\
\hline
First Step (4+10) 
& Both  & \textbf{0.9133} & \textbf{0.9230} & \textbf{0.9215} \\

\hline
\multirow{2}{*}{Second Step (7+7)} 
& Seed-based  & 0.8873 & 0.8965 & 0.8965 \\
& Augmentation  & \textbf{0.9085} & \textbf{0.9150} & \textbf{0.9158} \\

\hline
\multirow{2}{*}{Third Step (10+4)} 
& Seed-based  & 0.8741 & 0.8894 & 0.8900 \\
& Augmentation  & \textbf{0.8860} & \textbf{0.9021} & \textbf{0.9004} \\

\hline
\multirow{2}{*}{Final Step (13+1)} 
& Seed-based  & 0.8637 & 0.8847 & 0.8830 \\
& Augmentation  & \textbf{0.8852} & \textbf{0.8908} & \textbf{0.8960} \\
\hline
\end{tabular}
}
\end{table}

\subsubsection{Discussion}
The experimental results demonstrate that the proposed \emph{Incremental Attack Update Module} effectively balances adaptability to new attack categories with retention of prior knowledge. In the One‑Step Iteration setting, self‑training consistently achieves the highest performance across different values of $N$, indicating its strength in leveraging limited labeled data to generalize to unseen attacks. Active learning also performs competitively, particularly when the number of known classes increases, suggesting that uncertainty‑driven sample selection can mitigate label scarcity. In contrast, label propagation and label spreading show weaker performance, reflecting their sensitivity to noisy pseudo‑labels in heterogeneous attack distributions.

The Multi‑Step Iteration experiments highlight the benefits of progressive integration. Augmentation strategies, which incorporate pseudo‑labeled data into subsequent iterations, generally outperform strict seed‑based training, especially in intermediate steps (e.g., 7K+7U and 10K+4U). This indicates that incremental enrichment of the training set improves assimilation of new attack categories without severely compromising accuracy on previously known classes. However, seed‑based training demonstrates stronger stability in preserving performance on earlier classes, underscoring a trade‑off between knowledge retention and adaptability.

Overall, the results confirm three key properties of the proposed incremental learning pipeline:

\begin{itemize}
    \item \emph{Adaptability}: the system can continuously integrate new attack categories with minimal degradation.
    \item \emph{Retention}: previously learned knowledge remains stable, particularly under seed‑based training.
    \item \emph{Robustness under data scarcity}: high accuracy is maintained even when labeled data is limited, validating the effectiveness of semi‑supervised and active learning strategies.
\end{itemize}

These results suggest that combining self‑training with incremental augmentation in multi‑step updates provides a good balance between scalability and resilience, reinforcing the pipeline’s suitability for dynamic IIoT threat environments.

\section{\uppercase{LIMITATIONS}}
\label{sec:limitations}

There are several limitations to this study. First, all experiments were conducted using the Edge‑IIoTset dataset, which, while comprehensive, may not fully capture the diversity and evolving nature of real‑world IIoT traffic and attack behaviors. Second, the evaluation of the first DPM layer was restricted to three anomaly detection models and the second layer to four models, excluding other advanced or DL approaches that could provide different insights. Third, the novelty and outlier detection paradigms assume either completely clean training data or lightly contaminated environments, which may oversimplify the complexities of real deployment scenarios where contamination levels vary unpredictably. The incremental update experiments focused only on two incremental learning schemes, one-step and multi-step iterations, without exploring alternative paradigms such as continual lifelong learning or hybrid approaches. 

While these limitations exist, they are deliberate choices made to ensure the study remains focused, reproducible and computationally feasible. The exclusive use of the Edge‑IIoTset dataset provides a controlled and standardized benchmark, allowing for fair comparison across models and incremental learning schemes, even if it cannot fully capture the diversity of real‑world IIoT traffic. Similarly, restricting the evaluation to a small set of commonly used anomaly detection models enables a clear analysis of fundamental ML paradigms before extending to more complex DL approaches, which often introduce additional variables and resource demands. The assumptions in novelty and outlier detection paradigms, clean versus lightly contaminated training data, are widely adopted in anomaly detection research as they represent two fundamental scenarios that balance realism with feasibility. Finally, focusing on one‑step and multi‑step incremental learning schemes provides a structured foundation for evaluating adaptability and retention, while leaving continual lifelong learning and hybrid approaches as directions for future work.

\section{\uppercase{CONCLUSIONS}}
\label{sec:conclusions}

This study proposes MI$^2$DAS, a multi‑layer intrusion detection architecture for industrial IoT, designed to tackle important challenges such as high‑dimensional heterogeneous data, class imbalance, limited labeled samples and the emergence of zero‑day threats. The proposed MI$^2$DAS architecture integrates three layers that enable effective discrimination between normal and attack traffic, fine‑grained classification of attack types and continual learning to adapt to emerging threats. Using the Edge‑IIoT dataset, our experiments show that GMM achieves superior performance in normal‑attack separation at the edge (accuracy = 0.953, TPR = 1.000), while GMM and LOF complement each other in distinguishing known from unknown attacks. For known attack classification, the RF model outperformed alternatives, achieving a macro-F1 of 0.941 ± 0.054. Furthermore, semi‑supervised and active learning approaches effectively identified novel attacks with minimal labeling overhead, maintaining stable performance in both single‑step and multi‑step incremental updates. Future work will explore DL‑based anomaly detection methods to further improve recognition accuracy against complex and evolving attack patterns.

\bibliographystyle{apalike}
{\small
\bibliography{example}}



\end{document}